\documentclass[journal,onecolumn]{IEEEtran}

\usepackage{wrapfig}

\ifCLASSINFOpdf
\else
\fi
\usepackage{amsmath}

\def\code#1{\texttt{\small #1}}
\usepackage{graphicx}
\usepackage{xcolor}
\usepackage{listings}
\definecolor{strings_color}{rgb}{1.0,0.0,0}
\definecolor{comments_color}{rgb}{0.2,0.2,0.2}
 \usepackage{caption}
 \usepackage{chngcntr} 

\usepackage{hyperref}

\begin{document}
\title{NVIDIA Tensor Core Programmability, Performance \& Precision}


\author{\IEEEauthorblockN{Stefano Markidis, Steven Wei Der Chien, Erwin Laure}
\IEEEauthorblockA{\\\textit{KTH Royal Institute of Technology}}
\and
\IEEEauthorblockN{\\Ivy Bo Peng, Jeffrey S. Vetter}
\IEEEauthorblockA{\\\textit{Oak Ridge National Laboratory}}
}

\maketitle

\begin{abstract}
The NVIDIA Volta GPU microarchitecture introduces a specialized unit, called \emph{Tensor Core} that performs one matrix-multiply-and-accumulate on 4$\times$4 matrices per clock cycle. The NVIDIA Tesla V100 accelerator, featuring the Volta microarchitecture, provides 640 Tensor Cores with a theoretical peak performance of 125~Tflops/s in mixed precision. In this paper, we investigate current approaches to program NVIDIA Tensor Cores, their performances and the precision loss due to computation in mixed precision.

Currently, NVIDIA provides three different ways of programming matrix-multiply-and-accumulate on Tensor Cores: the CUDA Warp Matrix Multiply Accumulate (WMMA) API, CUTLASS, a templated library based on WMMA, and cuBLAS \code{GEMM}. After experimenting with different approaches, we found that NVIDIA Tensor Cores can deliver up to 83~Tflops/s in mixed precision on a Tesla V100 GPU, seven and three times the performance in single and half precision respectively.  A WMMA implementation of batched \code{GEMM} reaches a performance of 4~Tflops/s.  While precision loss due to matrix multiplication with half precision input might be critical in many HPC applications, it can be considerably reduced at the cost of increased computation. Our results indicate that HPC applications using matrix multiplications can strongly benefit from using of NVIDIA Tensor Cores.

\end{abstract}


\IEEEpeerreviewmaketitle

% First section: Intro
\section{Introduction}
The raising markets of AI-based data analytics and deep-learning applications, such as software for self-driving cars, have pushed several companies to develop specialized hardware to boost the performance of large dense matrix (tensor) computation. This is essential to both training and inferencing of deep learning applications~\cite{goodfellow2016deep}. For instance, Google designed the Tensor Processing Unit~\cite{jouppi2017datacenter} specifically for tensor calculations. Recently, NVIDIA released the Volta microarchitecture featuring specialized computing units called \emph{Tensor Cores}.

%that is specifically designed for tensor operations.
% Intel developed the Neural Network Processor~\cite{nnp}.  
%released in 2017\code{sgemm} 
%that have been the main workforce for training deep neural, 
%The new renaissance of neural networks is in part due to algorithm formulations in tensorial form that can take advantage of vector operations on Graphics Processing Units (GPU). 
%The NVIDIA Volta microarchitecture for first time features specialized Tensor Cores for increasing the performance of  deep-learning applications. 

An NVIDIA Tensor Core is capable of performing one matrix-multiply-and-accumulate operation on a 4$\times$4 matrix in one GPU clock cycle. In mixed-precision mode, Tensor Cores take input data in half floating-point precision, perform matrix multiplication in half precision and the accumulation in single precision. 

The NVIDIA Tesla V100 GPU provides a total of 640 Tensor Cores that can reach a theoretical peak performance of 125~Tflops/s. Hence, systems like the NVIDIA DGX-1 system that combines eight Tesla V100 GPUs could achieve a theoretical peak performance of one Pflops/s in mixed precision. The pre-exascale systems, such as the Summit supercomputer that has six Tesla V100 GPUs connected with high-speed NVLink in each compute node for a total of $~$4,600 nodes, will offer nearly 18M Tensor Cores!
% can provide immense computing power for HPC applications. ???
%This is a considerable amount of computing power in a single processing unit. 

%While calculations are in double precision, the majority of supercomputers in top500 November 2017 list have theoretical peak .

%However, it is important to understand which fraction of the peak performance it is possible to reach when using HPC workloads.

While large deep neural network applications will likely benefit from the use of NVIDIA Tensor Cores, it is still unclear how traditional HPC applications can exploit Tensor Cores. We envision three challenges. The first is to determine suitable programming models for Tensor Cores. In this work, we try to understand which programming interface can provide rich expressiveness while enabling maximum performance. The second is to quantify performance improvement from using Tensor Cores for various problem sizes and workloads. The third is to quantify the loss of precision when using mixed precision operations and to design techniques to improve the accuracy. We foresee these challenges will be of paramount importance for the HPC community as incoming supercomputers, such as Sierra and Summit, will be equipped with NVIDIA Tesla V100 GPUs. HPC applications running on these systems will need to take advantage of the NVIDIA Tensor Cores to reach maximum performance.
% In this paper, we provide timely experiments that identify the crossover when the usage Tensor Cores becomes advantageous. 
%Ee propose a novel technique in this study for reducing precision loss with low computation cost.

The objective of this paper is to evaluate the three challenges by providing an up-to-date study of NVIDIA Tensor Core. In particular, we focus on programmability, performance and precision loss in the context of HPC applications. We summarize our main contributions as follows:
\begin{itemize}
\item We survey current programming interfaces that perform tensor operations on NVIDIA Tensor Cores.
\item We characterize the performance of NVIDIA Tensor Cores when computing large matrix multiplication and batched matrix multiplications. We compare them with the performance of the same operations on CUDA cores to quantify the performance boost.
\item We quantify precision loss in matrix multiplication due to half precision matrix input with varying matrix sizes.
\item We propose a technique to decrease precision loss in matrix multiplications on Tensor Cores at the cost of increased computation.
\end{itemize}

The paper is organized as follows. We first introduce previous works related to tensor architectures in Section~\ref{relatedwork}. We briefly describe NVIDIA Volta microarchitecture in Section~\ref{VoltaArchitecture}.  We then present current approaches for programming NVIDIA Tensor Cores in Section~\ref{programming}.  We focus on precision loss due to NVIDIA Tensor Core mixed-precision and possible methods to decrease precision loss in Section~\ref{precision}. We describe our experimental set-up in Section~\ref{experiments} and present the performance results in Section~\ref{results}. Finally, Section~\ref{discussion} discusses the main results of this work and concludes the paper.

\section{Related Work}
\label{relatedwork}
AI-based data analytics and deep neural network applications have become increasingly important in recent years. These applications lead to rapid development of software and hardware that efficiently express and support tensor operations, which are fundamental for deep neural network applications. TensorFlow is among the most popular open-source programming framework that uses a computational graph with tensor operations as nodes of the graph~\cite{abadi2016tensorflow}. Caffe,  Torch and Microsoft CNTK are other popular programming frameworks for developing deep neural networks~\cite{goodfellow2016deep}.

The seminal paper by Krizhevsky \emph{et al.}~\cite{krizhevsky2012imagenet} has established GPUs as the main workforce in training deep neural networks and triggered a renaissance of deep-learning applications. Besides NVIDIA Tensor Cores~\cite{whitepaper2017} discussed in this paper, several companies are also employing and developing specialized hardware for high-performance inference. Microsoft deployed the Catapult system that uses FPGAs~\cite{putnam2014reconfigurable}. Movidius developed the Myriad 2 Vision Processing Unit~\cite{barry2015always}. Google designed and developed Tensor Processing Unit (TPU) specifically for inference workloads. The main engine of the TPU is a MAC matrix multiply unit containing 256$\times$256 MACs, each capable of performing 8-bit multiply-and-adds on signed or unsigned integers. In December 2017, Intel announced the release of the Neural Network Processor (NPP)~\cite{nnp}, which implements a new memory architecture for tensor operations. NPP does not have standard caches and data movement is programmable by software. In addition, neuromorphic hardware, such as the IBM TrueNorth~\cite{merolla2014million} and SpiNNaker~\cite{khan2008spinnaker} chips, mimics the functioning of spiking neural network. Although their original design purpose is to simulate the brain, they may also find usage in AI applications.

These new architectures usually have lower power and energy footprint than general processors that are employed in traditional HPC systems. The reduced power consumption mainly comes from reduced accuracy in computation by using fewer bits for representation. In deep neural networks, floating-points are transformed to narrow precision via quantization. For instance, TPU operates on eight-bit integers. NVIDIA Tensor Cores follows the IEEE 754 standard~\cite{whitehead2011precision} and uses mixed floating-point precision, i.e., matrix multiplication input in half precision and accumulation in single precision. Intel NPP introduces a new format, called \emph{Flexpoint}~\cite{koster2017flexpoint}. Flexpoint uses fixed-point multiplications and a shared exponent to allow a large dynamic range. While several studies have shown that deep neural networks are tolerant to low precision calculation~\cite{micikevicius2017mixed, gupta2015deep, courbariaux2014low}, such studies are still in their infancies in HPC. Mixed single and double precision calculations have been studied in the context of HPC~\cite{buttari2007mixed, haidar2017investigating}. However, these emerging architectures have rather narrow precision, smaller than single precision, and the topic is still to be studied in details. 

%  Especially in the HPC context.
% One of the goals of this paper is to understand the precision loss with NVIDIA Tensor Cores and develop a method to decrease the precision loss.

\section{NVIDIA Volta Architecture}

\label{VoltaArchitecture}
In May 2017, NVIDIA released Volta GV100 GPU architecture and the Tesla V100 accelerator to boost AI and HPC applications. As one of the largest silicon chips, Volta GPU includes 21.1 billion transistors on a die area of 815~mm\textsuperscript{2}. 

\begin{figure}[h]%
	\begin{center}
		\includegraphics[width=0.6\columnwidth]{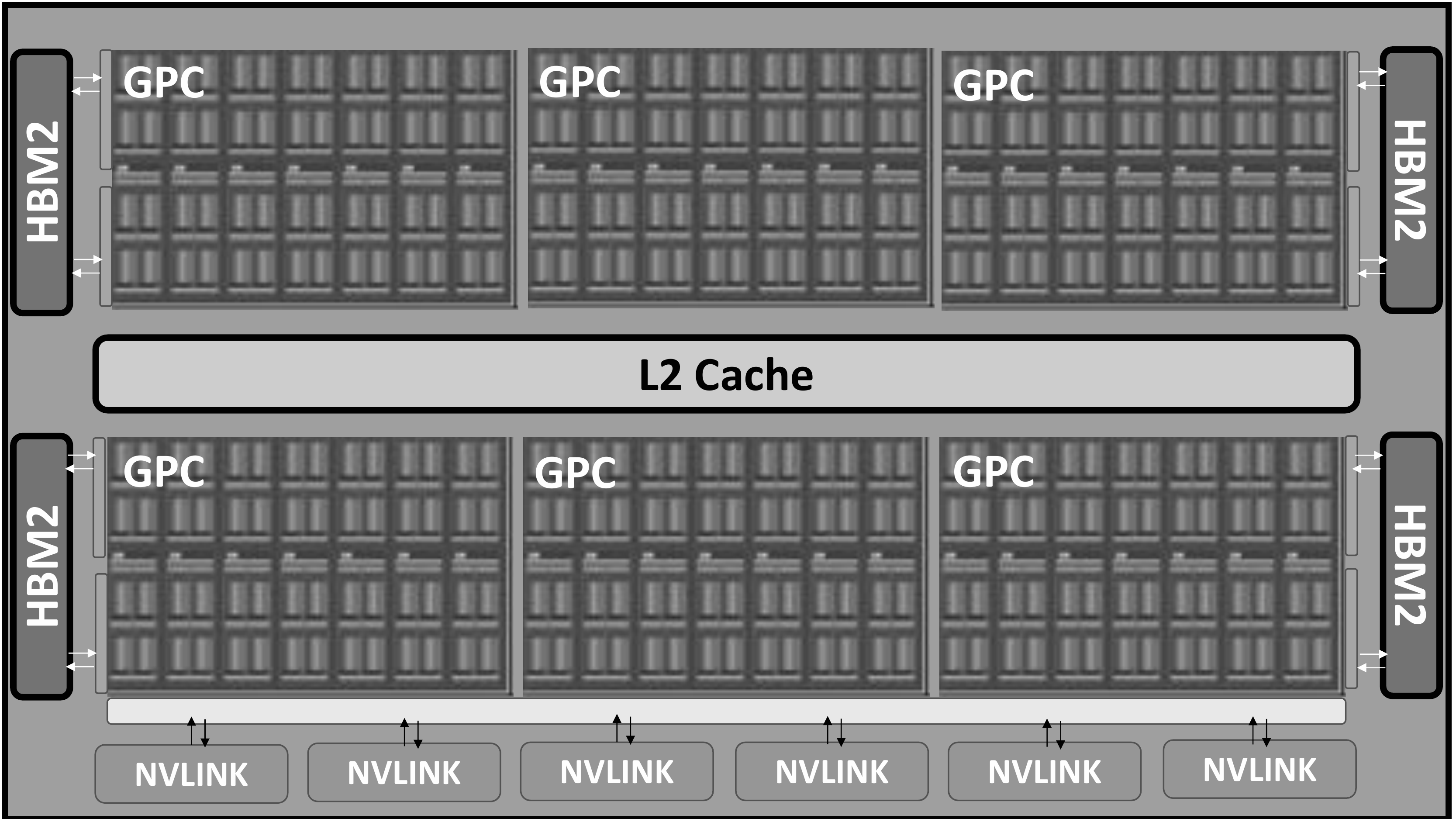}
		\caption{Volta GV100 GPU architecture features six GPCs and 16 GB HBM2. Adapted from~\cite{durant2017inside}.}%
		\label{gv100}
	\end{center}
\end{figure}

A full GV100 GPU consists of six GPU Processing Clusters (GPCs). Each GPC contains seven Texture Processing Clusters (TPCs) and 14 Streaming Multiprocessors (SMs). A 16 GB HBM2 memory, connecting through eight memory controllers in four memory stacks, is embedded in the same package. We present the architecture of GV100 GPU in the simplified diagram of Fig.~\ref{gv100}.

\begin{figure}[h]%
\begin{center}
\includegraphics[width=0.6\columnwidth]{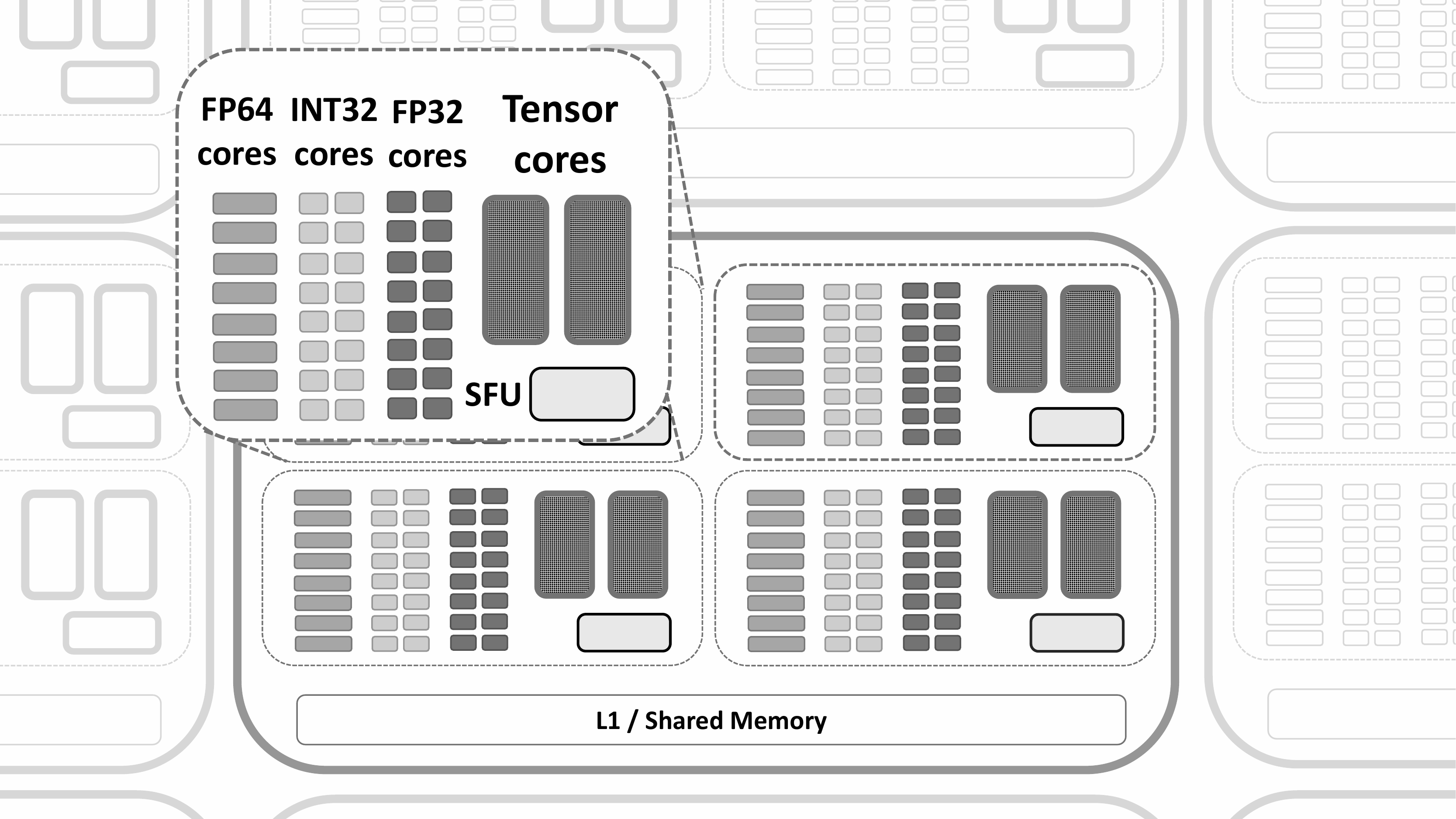}
\caption{Simplified diagram of the Volta SM architecture. The NVIDIA Tesla V100 uses 80 SMs.\label{Architecture}}
\end{center}
\end{figure}

The Volta microarchitecture features a renewed Streaming Multiprocessor (SM) design~\cite{durant2017inside} (Figure~\ref{Architecture}). Each SM is partitioned into four processing blocks. Each block consists of two Tensor Cores, 8 FP64 cores, 16 FP32 cores, 16 INT32 cores and one Special Function Unit (SFU). One main design change in Volta SM is the integration of L1 data cache and shared memory subsystem. Their combined capacity of 128 KB per SM is 7$\times$ larger than the data cache of Volta's predecessor GP100 GPU. Also, texture units and SMs can share this merged L1 cache/shared memory and configure up to 96 KB shared memory per SM. The Tesla V100 accelerator uses 80 SMs for a total of 2,560 FP64 cores, 5,120 FP32 cores and 640 Tensor Cores.

\begin{figure}
\begin{center}
\includegraphics[width=0.6\textwidth]{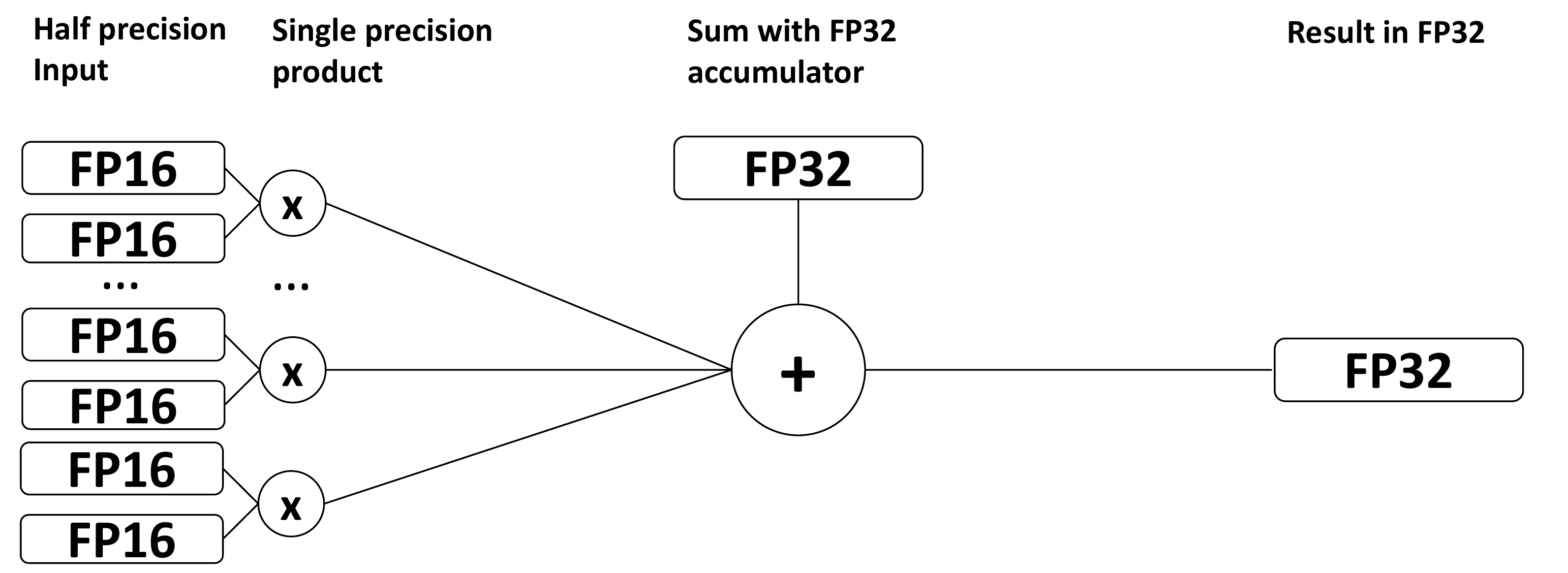}
\caption{FMAs in NVIDIA Tensor Cores.\label{FMA}}
\end{center}
\end{figure}

A new feature of Volta SM is mixed-precision operations with Tensor Cores. In each cycle, a Tensor Core can perform 64 floating-point Fused-Multiply-Add (FMA) operations~\cite{whitepaper2017}. An FMA operation takes input values in half precision while the output values can be either in half (FP16) or full precision (FP32) as illustrated in Fig.~\ref{FMA}. FMA has the advantage of using only one rounding operation instead of two, resulting in a more accurate output~\cite{whitehead2011precision}.

In total, the Tesla V100 accelerator can perform up to 40,960 FMA operations per cycle, i.e., 81,920 floating-point operations. The Tesla V100 accelerator uses the base clock frequency 1.3~GHz and it can be boosted to 1.53~GHz. The theoretical maximum performance can reach 31.4~Tflops/s with half precision, 15.7~Tflops/s with single precision, and 7.8~Tflops/s with double precision while  Tensor Cores can deliver 125~Tflops/s.

Half precision data and instruction in Tensor Cores are the main contributors to high throughput. Compared to single precision, half precision data only requires half memory bandwidth and footprint, resulting in faster data transfer. 

%Obviously, this performance optimization is a tradeoff of accuracy. 
%Thus, half precision is mostly beneficial for training neural network where some back-propagation algorithms are resilient to loss of precision. 
%By design, using higher precision (FP32) in Tensor Cores for accumulation in FMA operations could help mitigate precision loss in some deep-learning applications. 
%Additionalyalso major new software innovations are introduced.
%\subsection{Nvidia Tensor Cores}

\section{Programming NVIDIA Tensor Cores}
\label{programming}
The NVIDIA Tensor Core basically performs only one kind of operation: matrix-multiply-and-accumulate on 4$\times$4 matrices. Therefore, a programming interface for NVIDIA Tensor Cores can simply express the BLAS \code{GEMM} (GEneral Matrix to Matrix Multiplication) operation. A \code{GEMM} operation consists of the multiplication of two matrices $A$ and $B$ and accumulation of the result into a third matrix $C$, i.e. $C = \alpha AB + \beta C$. Here we present different interfaces of Tensor Cores to illustrate their programmability with different levels of abstraction.
% with $\alpha=1$ and $\beta=1$.

% changes made to address reviewer's suggestion on storyline

Currently, the lowest level interface to program NVIDIA Tensor Cores is CUDA~9 Warp Matrix Multiply and Accumulation (WMMA) API. CUDA~9 WMMA is a CUDA preview feature and WMMA will likely be changed in future releases with no backward compatibility guarantee. We briefly present it as at the moment it is the only way to program Tensor Cores directly and future APIs might be developed upon CUDA~9 WMMA.

CUDA~9 allows us to program a basic matrix-multiply-and-accumulate on 16$\times$16 matrices. Recent CUDA~9 releases, such as CUDA~9.1, also support non-square matrix multiplication with different sizes. We note that while NVIDIA Tensor Core implements 4$\times$4 matrix multiplications in hardware, CUDA~9 WMMA allows us only to compute larger matrix multiplications. This is in-line with the CUDA philosophy of running many more threads than hardware computing units (problem over-decomposition) to hide instruction and memory latencies.

%First, the two input matrices, \code{A} and \code{B}, in half precision are passed to the kernel and loaded into fragments \code{Amat} and \code{Bmat}. \code{A} and \code{Amat} have size M$\times$K while \code{B} and \code{Bmat} have size K$\times$N. In the Listing~\ref{exampleCode} M, K, and D are defined as macro variables. In our experiments, we use CUDA~9.0 which supports only M = K = N = 16.

%CUDA 9.1 also supports non-square matrix multiplications with M = 32, K =  8,  N = 16 and M = 8, K =  32,  N = 16.
\begin{lstlisting}[caption={CUDA~9 WMMA provides a direct way to calculate 16x16 matrix matrix-multiply-and-accumulate using a CUDA Warp (32 threads).}, label={exampleCode}]
// Calculate AB with NVIDIA Tensor Cores 
// Kernel executed by 1 Warp (32 Threads)
__global__ void tensorOp(float *D,half *A,half *B) {
	// 1. Declare the fragments
	wmma::fragment<wmma::matrix_a, M, N, K, half, wmma::col_major> Amat;
	wmma::fragment<wmma::matrix_b, M, N, K, half, wmma::col_major> Bmat;
	wmma::fragment<wmma::accumulator, M, N, K, float, void> Cmat;
	// 2. Initialize the output to zero
	wmma::fill_fragment(Cmat, 0.0f);
	// 3. Load the inputs into the fragments
	wmma::load_matrix_sync(Amat, A, M); 
	wmma::load_matrix_sync(Bmat, B, K); 
	// 4. Perform the matrix multiplication 
	wmma::mma_sync(Cmat, Amat, Bmat, Cmat);
	// 5. Store the result from fragment to global
	wmma::store_matrix_sync(D,Cmat, M, wmma::mem_col_major);
}
\end{lstlisting}

Listing \ref{exampleCode} presents a CUDA kernel that performs a matrix multiplication of two 16$\times$16 matrices with one CUDA Warp (32 threads). The kernel consists of five parts. First, the WMMA fragments (GPU register memory for storing the input matrices) \code{Amat}, \code{Bmat} and \code{Cmat} are declared. Second, the accumulator fragment, \code{Cmat}, for storing the result of the matrix multiply, is set to zero. Third, the input matrices are loaded into the fragments \code{Amat}, \code{Bmat} using \code{wmma::load\_matrix\_sync()}. Fourth, the multiplication is performed by calling the \code{wmma::mma\_sync()}. Finally, we move the results from the fragment \code{Cmat} to \code{D} in the GPU global memory. Each matrix multiplication and accumulation should be executed by one CUDA Warp (32 threads). If the kernel \code{tensorOp} is launched with less than 32 threads, the result of the matrix multiplication is undetermined. On the other hand, using more threads than a Warp will still result in the correct results.

An important point is that the two-dimensional tensors are provided as 1-D arrays. For this reason, we need to declare if the 1-D arrays should be interpreted either as row- or column-major.

\subsection{Matrix Multiplication}
While CUDA~9 WMMA provides a direct way of performing \code{GEMM} only with fixed-size matrices, three other methods can be used to calculate matrix multiplications of arbitrary size:
\begin{itemize}
\item {\bf Tiled Matrix Multiply with CUDA~9 WMMA.} With this technique, the result matrix, $C$ is divided in fixed-size tiles (sub-matrices), i.e. 16$\times$16, and each of the $C$ tile values can be calculated by summing the result of $A$ and $B$ tile multiplications. This tiling technique for matrix multiplication is widely used in GPU programming to exploit \code{shared} GPU memory. One thread block per tile is used~\cite{kirk2016programming} while in the case of CUDA~9 WMMA, a Warp is assigned to the tile.
\item {\bf NVIDIA CUTLASS} (CUDA Templates for Linear Algebra Subroutines) is a CUDA C++ templated header-only library to perform \code{GEMM} operation in different precisions (\code{dgemm}, \code{sgemm} and \code{hgemm})~\cite{CUTLASSref}. It supports also CUDA~9 WMMA implementation (\code{wgemm}). The library supports different tiling strategies and exploits software pipelining to hide GPU memory latencies.
\item {\bf NVIDIA cuBLAS} is an NVIDIA library that implements standard basic linear algebra subroutines (BLAS)~\cite{nvidia2008cublas}. The library provides \code{GEMM} routines for Tensor Cores. In order to perform \code{GEMM} on NVIDIA tensor Cores, the cuBLAS math mode needs to be set to \code{CUBLAS\_tensorOp\_MATH} using the function \code{cublasSetMathMode()}. It is then possible to use either \code{cublasGemmEx()} or \code{cublasSgemm()} to perform \code{GEMM} on NVIDIA tensor Cores.
\end{itemize}
%In addition, cuDNN also support convolution using images. In this paper, we focus on HPC aspects of Tensor Cores and for this reason, we didn't investigate the performance of convolution of images.

\subsection{Batched Matrix Multiplications}
Many HPC applications rely on the solution of several small-size matrix multiplications in parallel~\cite{dongarra2017495}. One example is the Nek5000 CFD application that uses small-size matrix multiplies for each spectral element resulting from the semi-spectral discretization~\cite{offermans2016strong,markidis2015openacc}. In this case, the matrix size depends on the order of the spectral element in each direction. Another application is the Fast Multipole Method-accelerated Fast Fourier Transform (FFT) that requires also many small matrix multiplications~\cite{cecka2017low}. BLAS \code{GEMM} routines are optimized for solving large matrix multiplications and do not perform optimally in solving small-size matrix multiplications. Libraries, such as LIBXSMM~\cite{heinecke2016libxsmm} and Intel MKL, provide high-performance small-size matrix multiplications.

The most convenient approach to solve several small matrix multiplication in parallel on GPU is through NVIDIA cuBLAS. The NVIDIA cuBLAS library provides a batched \code{sgemm} API for single precision matrix multiply, called \code{cublasSgemmBatched()}. However, batched \code{GEMM} is not supported by NVIDIA Tensor Cores\footnote{\label{new-cublas-release}After the completion of this work, batched \code{GEMM} API for Tensor Cores was released in cuBLAS 9.1.128, among other optimizations~\cite{cuda-release-notes}.}. In this work, we implement a simple batched \code{GEMM}, based on Listing~\ref{exampleCode}, to evaluate the possible performance benefit of using NVIDIA Tensor cores to solve batched \code{GEMM}.

\section{Precision Loss}
\label{precision}
Each Tensor Core performs a multiplication of two matrices with half precision floating-point entries and adds the result to an accumulator in single precision (see Fig.~\ref{FMA}). The use of mixed precision calculations might cause large rounding errors, affecting simulation accuracy.

One of the motivations for matrix multiplication in half precision is that the matrix entries that are multiplied in neural network are small with respect to the value of the previous iteration. For this reason, the multiplication result is still small in value. However, the result is accumulated to another value that might be much larger. To avoid precision loss or use additional computation, i.e. Kahan summation~\cite{higham1993accuracy}, accumulation is performed in single precision. 

\begin{figure}[h]%
\begin{center}
\includegraphics[width=0.6\columnwidth]{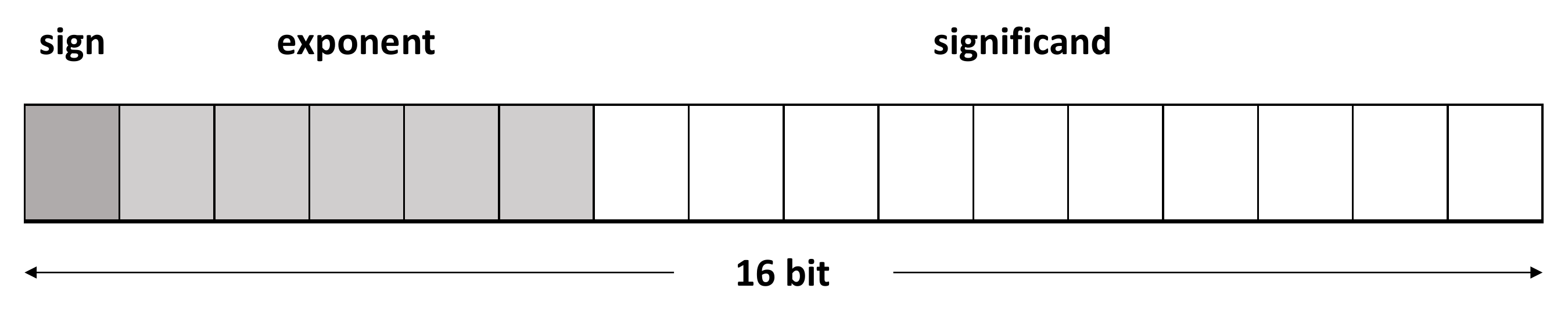}
\caption{Half precision floating-point number representation.\label{halfprecision}}
\end{center}
\end{figure}

In addition, deep neural network training are tolerant to precision loss up to certain degree~\cite{micikevicius2017mixed, gupta2015deep, courbariaux2014low}. Thus, high precision calculations are not critical for the completion of many deep neural network trainings. On the other hand, the vast majority of traditional HPC applications, with probably the exception of Montecarlo codes, are considerably more sensitive to rounding errors that arise from the usage of narrow precision. For this reason, it is important to characterize the impact of mixed precision calculations in widely used HPC computational kernels, such as \code{GEMM}. Narrow precision matrix multiplications might severely impact the possible usage of NVIDA Tensor Cores in HPC applications. Half precision floating-point representation uses 16 bits: one bit for the sign, five bit for the exponent and ten bits for the significand (or fraction or mantissa), as illustrated in Fig.~\ref{halfprecision}. 

The limited number of bits of floating-point number representation introduces two limitations:
\begin{itemize}
\item {\bf Limited range.} Having only five bits of exponent, the maximum representable number in half precision is 65,504 and the range of half precision floating-point is $\pm$65,504. When rounding a value from single to half precision, if the float number is larger than 65,504, it is set to half infinity.  The machine epsilon in half precision floating numbers is $2^{-10}$. Any float number that is too small to be represented as a half will be set to zero.
\item {\bf Decreasing precision with increasing value range intervals.}  The most striking characteristic of using half precision is the extreme precision loss for large numbers. In fact, when using half precision, we have only 1,024 values for each power of two number intervals. For instance, there are 1,024 representable numbers between one ($2^0$) and two ($2^1$). Between 1,024 ($2^{10}$) and 2048 ($2^{11}$), there are also only 1,024 values so all the fractional precision is lost for numbers larger than 1,024. For the same reason, there is only an accuracy of $\pm$32 between 32,768 ($2^{15}$) and 65,536 ($2^{16}$). 
\end{itemize}
It is clear from this brief discussion that precision strongly depends on the value range of the numbers used in simulations: rounding relatively large number from single to half precision leads to considerable precision loss.

While the impact of half precision input on large HPC applications requires in-depth studies~\cite{luszczek2017towards}, we observe that it is possible to decrease the precision loss in matrix multiplications, $C = AB$, at the cost of increased computation and memory consumption with a simple technique. 

We define a half-precision residual matrix $R$ as the difference between a matrix before and after rounding from single to half-precision, where $A_{single}$ and $A_{half}$ represents a matrix before and after rounding (notation is analogous with matrix $B$):
\begin{equation}
R_{A} = A_{single} - A_{half}.
\end{equation}
We manipulate $A_{single}B_{half}$ and compute on Tensor Cores using the distributive property of matrix multiplication and sum as
\begin{equation}
\begin{split}
A_{single}B_{half} = (A_{single} - A_{half} + A_{half})B_{half} = (R_{A} + A_{half})B_{half} =  R_{A}B_{half}  +  A_{half} B_{half}.
\end{split}
\label{eq1}
\end{equation}

The equation above allows us to take into account the rounding error from single to half-precision for matrix $A$ with one additional matrix multiplication on NVIDIA Tensor Cores and additional memory for storing $R_{A}$. We call this simple technique \emph{precision refinement}, as it is similar to analogous techniques, called \emph{iterative precision refinement} and used in other works for the solution of linear systems~\cite{luszczek2017towards}.

Since $B_{half}$ is still rounded directly from $B_{single}$, precision loss is only partially eliminated. It is possible to further recover precision by applying the same technique again to matrix $B$ with $R_{B} = B_{single} - B_{half}$ and apply the distributive property of matrix multiplication and sum:
\begin{equation}
\begin{split}
A_{single}B_{single}  = (R_{A} + A_{half})(R_{B} + B_{half}) = R_{A}R_{B}  + A_{half}R_{B}  + R_{A} B_{half}  +  A_{half} B_{half}.
\end{split}
\label{eq2}
\end{equation}
In this case, we can reduce the precision loss by performing four matrix multiplications on the NVIDIA Tensor Cores and using additional memory for storing $R_{A}$ and $R_{B}$.

We motivate our method by the assumption that the precision loss due to conversion arises from the fact that 16-bit cannot entirely represent all values in 32-bit. Thus, we distribute the un-representable portion of the value (residual) to another 16-bit number. Since the value is originally in 32-bit, it can be fully represented by two 16-bit numbers, subject to error from distribution. In other words, we recover the loss in precision due to input conversion by additional operation with residual values that were recorded during conversion. With this scheme, depending on the precision requirement of an application, the developer can choose to perform refinement on one or both matrices at the expense of additional computation time and memory.
 %We explain the increase in error with increasing dimension, even when four matrix multiplications are used, by that error from distribution accumulates during the multiplication process.

%\begin{lstlisting}[caption={Implementation of refinement by accumulating final result as matrix C with \code{cublasGemmEx}}, label={refinement}]
%float zero = 0.0f, one = 1.0f, alpha = 1.0f;
%cublasGemmEx(...,&alpha,d_A_half,CUDA_R_16F,M,
%	d_B_half,CUDA_R_16F,K,
%	&zero,d_C_refined,CUDA_R_32F,M,
%	CUDA_R_32F,CUBLAS_GEMM_DEFAULT_TENSOR_OP));
%	
%cublasGemmEx(...,&alpha,d_A_r,CUDA_R_16F,M,
%	d_B_half, CUDA_R_16F, K,
%	&one,d_C_refined, CUDA_R_32F, M,
%	CUDA_R_32F, CUBLAS_GEMM_DEFAULT_TENSOR_OP));
%	
%cublasGemmEx(...,&alpha,d_A_half, CUDA_R_16F,M,
%	d_B_r, CUDA_R_16F, K,
%	&one,d_C_refined, CUDA_R_32F, M,
%	CUDA_R_32F, CUBLAS_GEMM_DEFAULT_TENSOR_OP));
%	
%cublasGemmEx(..., &alpha,d_A_r,CUDA_R_16F,M,
%	d_B_r,CUDA_R_16F,K,
%	&one,d_C_double_tuned,CUDA_R_32F,M,
%	CUDA_R_32F,CUBLAS_GEMM_DEFAULT_TENSOR_OP));
%\end{lstlisting}
%

%%%% EXPERIMENTAL SETUP
\section{Experimental Set-up}
\label{experiments}
We test NVIDIA Tensor Cores with a Tesla V100 accelerator which is connected to an Intel E5-2690v3 Haswell host. The Operating System is CentOS Linux version 7.4.1708. We use CUDA Driver/Runtime Version 9.0 with CUDA Capability 7.0. The GNU compiler version for compiling host code is 4.8.5. The \code{nvcc} compiler flags \code{-O3 -Xptxas -v -std=c++11 -gencode arch=compute\_70,code=sm\_70 -gencode arch=compute\_70,code=compute\_70} are used. The tested Tesla V100 supports a base default GPU clock at 1.245~GHz and a boost GPU clock at 1.38~GHz. In this paper, we report the results using the boost GPU clock at 1.38~GHz. We note that the GPU boost frequency in our system is 10\% lower than the GPU boost frequency reported in Ref.~\cite{durant2017inside}. With GPU clock at 1.38~GHz, the theoretical peak performance on Tensor Cores is 112.7~Tflops/s. 

We measure performance of Tensor Cores using \code{GEMM} operation, $C = \alpha AB + \beta C$ with $\alpha=1.0$ and $\beta=1.0$. We initialize $A$, $B$ and $C$ values in single floating-point precision. When the \code{GEMM} is computed on the Tensor Cores, the values of $A$ and $B$ are first rounded to half precision. The time to complete the rounding is not considered when timing the overall execution time. 

We report the results using square matrices with size $N$ for each dimension. We take Tflops/s as the main figure of merit for performance. To time the CUDA execution of kernels running on the GPU, we use CUDA events that have a resolution of approximately half microsecond.  The number of operations are calculated assuming that the matrix multiplication uses the naive algorithm requiring $\mathcal{O}(N^3)$ operations. We note that cuBLAS \code{GEMM} matrix multiplication might use other matrix multiplication algorithms, i.e. Strassen algorithm. In this case, the performance of the cuBLAS might be affected by the algorithm in use also.

We run 5 to 100 tests and present the harmonic mean of flops/s in the plots. If the execution time is taken as the performance figure of merit, we report the arithmetic mean of execution times. We do not show error bars when the error is less than 1\%.

For the sake of comparison, we also report the performance of a naive implementation using CUDA~9 WMMA without any optimization (see Listing~\ref{exampleCode}), such as the use of CUDA \code{shared} memory and software pipeline. CUTLASS makes use of these techniques to provide an optimized use of CUDA~9 WMMA. When we measure the performance of CUTLASS, we tested different tiling techniques with different execution configurations; we report the timing of the set-up with higher performance for a given $N$. 

As NVIDIA does not provide yet a batched \code{GEMM} for Tensor Cores~\textsuperscript{\ref{new-cublas-release}}, we wrote a simple implementation for testing purposes, extending the code in Listing~\ref{exampleCode}. For batched \code{GEMM}, we only use square 16$\times$16 matrices. In this case, the CUDA execution configuration consists of 512 threads per block. Since a 16$\times$16 matrix multiplication is executed by one Warp (32 threads), 16 matrix multiplications are executed per thread block.

\begin{figure}[t]
\begin{center}
\includegraphics[width=0.6\columnwidth]{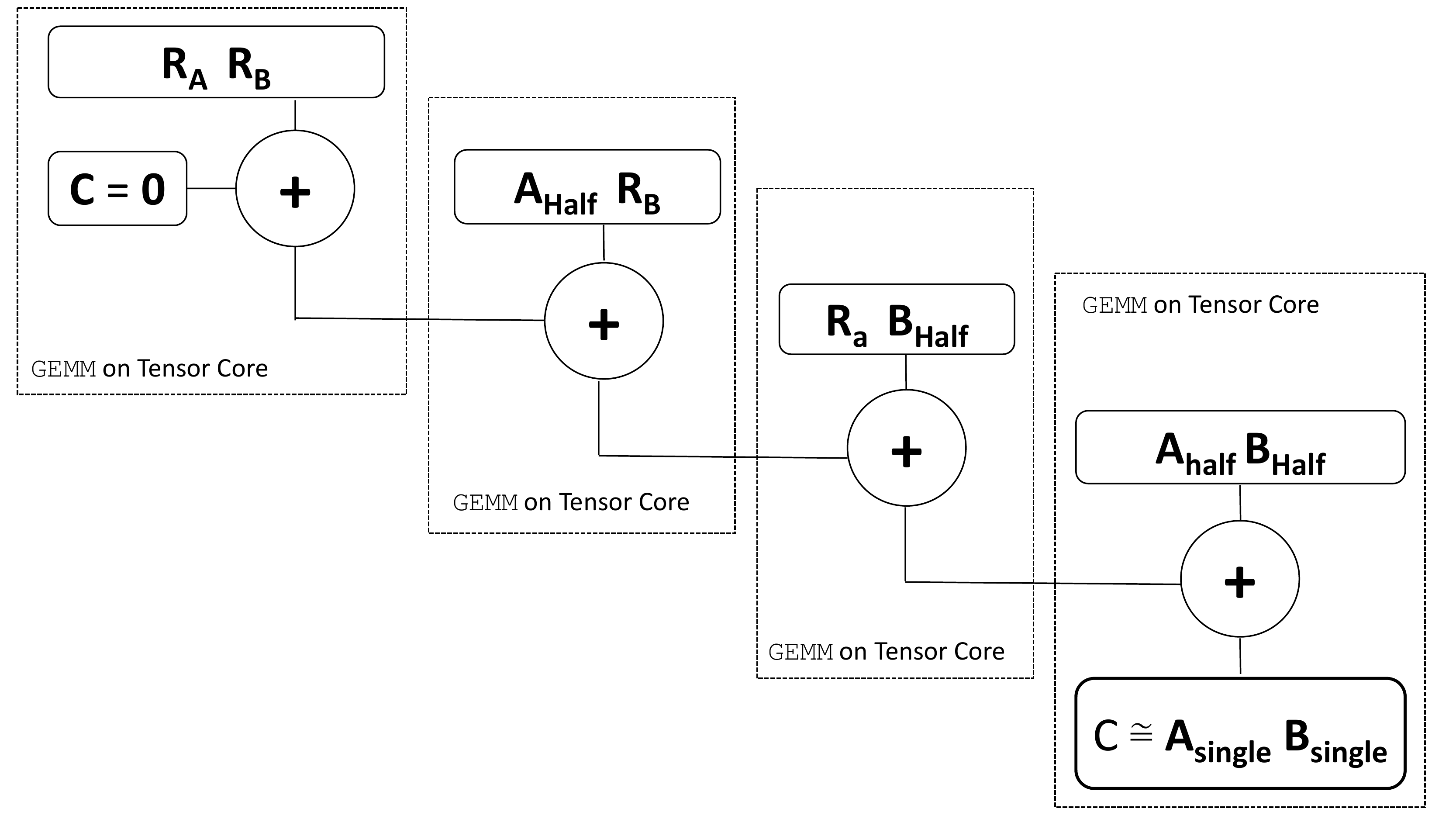}
\caption{Implementation of precision refinement using four pipelined \code{GEMM} on Tensor Cores.}% with $\beta = 0$ 
\label{pipeline}%
\end{center}
\end{figure}

To quantify the precision loss due to mixed precision computations, we first calculate the error matrix $e$ as $e = (C_{half}  - C_{single})$, where $C_{half}$ is the result of the matrix multiplication with half precision input and $C_{single}$ is the result with single precision input. We then apply the max norm $\Vert e \Vert_{Max} = \max(|e_{i,j}|)$. We choose the max norm to quantify the error as it provides a bound of the maximum error per matrix entry. We initialize the two square matrices $A$ and $B$ of size $N$ with random numbers, taken from range [-1,1] in single precision. The matrix values are then converted to half precision. We then vary the matrix size $N$ to study how the total number of operations affects the overall precision loss.

In addition, we implement Eq.~\ref{eq1} (precision refinement with $R_A$) and Eq.~\ref{eq2} (precision refinement with both $R_A$ and $R_B$) to assess the computational cost of techniques to reduce the precision loss. The diagram in Fig.~\ref{pipeline} shows the implementation of Eq.~\ref{eq2} using four pipelined \code{GEMM} to perform matrix multiplications on Tensor Cores. In this case, we use a quick implementation based on four cuBLAS function calls such that the result of a \code{GEMM} is used as half precision input for the next \code{GEMM}. We note that optimized versions of such techniques are possible. We provide a simple implementation for fast comparison and estimation of the computational cost for decreasing the precision loss. %with $\beta = 0$ 

% and Since our scheme requires four matrix-matrix multiplication and three matrix addition. We implemented the scheme with a simple pipelining technique by manipulating the value of Beta in \code{cublasSgemmEx} API. We leverage the fact that accumulation is done in full-precision and accumulate our final result in matrix C by setting beta to zero initially and to one for operations thereafter. 

%When comparing the performance of \code{sgemm} with and without Tensor Cores, it is important to note that \code{sgemm} is run in fully single and half precision on CUDA cores and mixed precision on NVIDIA Tensor Cores. So the performance tests we report in Section \ref{results} exploit computation in different precisions. 

\section{Results}
\label{results}
\begin{figure}[t!]%
\begin{center}
\includegraphics[width=0.6\columnwidth]{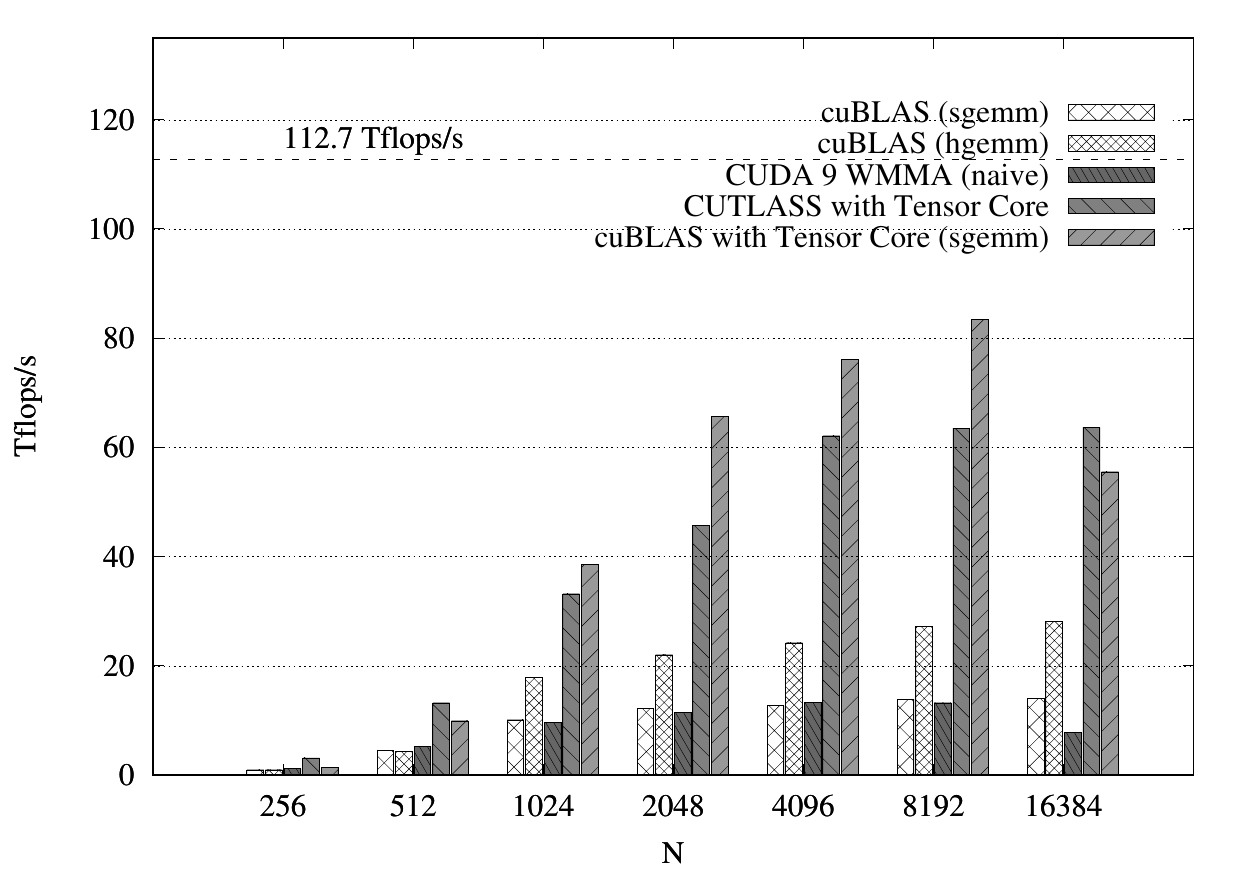}
\caption{\code{GEMM} performance without Tensor Cores in single and half precision (white bars) and with Tensor Cores using naive implementation with CUDA 9 WMMA, CUTLASS and cuBLAS (grey bars) varying with matrix size $N$.}%
\label{performance-sgemm}%
\end{center}
\end{figure}

In this section, we present and discuss the experimental results. Our results show that using NVIDIA Tensor Cores to compute \code{GEMM} can lead to considerable performance boost. Fig.~\ref{performance-sgemm} presents the \code{GEMM} performance with and without Tensor Cores. The bars in white color show the \code{GEMM} performance with CUDA cores in full single and half precision without Tensor Cores. The bars in grey color show the \code{GEMM} performance on Tensor Cores using a naive implementation with CUDA~9 WMMA, CUTLASS and cuBLAS respectively. In addition, a line at 112.7~Tflops/s  (theoretical peak using Tensor Cores in our system) is superimposed to the plot.

\subsection{Performance}

We achieved maximum performance of 83~Tflops/s on NVIDIA Tensor Cores for $N=8,192$ using cuBLAS \code{GEMM}. The measured peak performance in mixed precision is approximately 74\% the theoretical performance of the NVIDIA Tensor Cores, which is about 6$\times$ and 3$\times$ the performance of \code{GEMM} in full single and half precision. For $N=16,384$, CUTLASS performs better than cuBLAS \code{GEMM} on Tensor Cores. This is probably due to the fact that CUTLASS can be tested with different tiling configurations to select the most performant setup.

The naive CUDA~9 WMMA implementation does not provide any performance improvement with respect to \code{sgemm} on the CUDA cores. Also, it is outperformed by the \code{hgemm} in half precision. If the \code{GEMM} implementation with CUDA~9 WMMA also includes the use of CUDA \code{shared} memory, the performance (not shown here) is about five times higher than the performance of the naive implementation for $N=8,192$. This indicates that it is critical to use CUDA \code{shared} memory to reduce memory traffic \cite{kirk2016programming} when programming NVIDIA Tensor Cores.

\begin{figure}[t!]%
\begin{center}
\includegraphics[width=0.6\columnwidth]{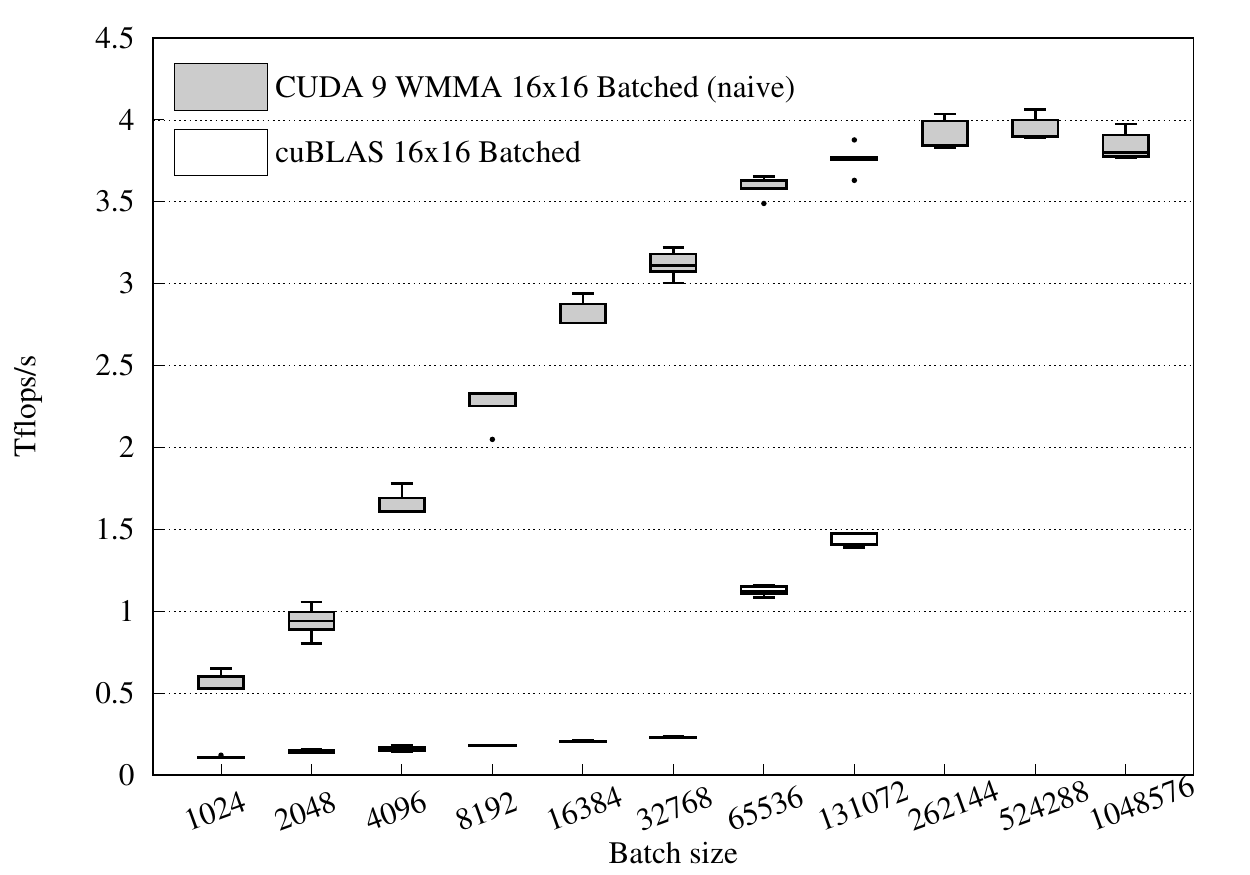}
\caption{Performance of cuBLAS batched \code{sgemm} on CUDA cores, and CUDA~9 WMMA implementation performing \emph{batch size} 16$\times$16 matrix multiplies. The cuBLAS batched \code{sgemm} cannot run for more than 131,072 multiplications as they require more memory than the available one on the Tesla V100 GPU.}%
\label{performance-batched}%
\end{center}
\end{figure}

We also evaluate the potential performance improvement when running batched \code{GEMM} on Tensor Cores. We compare the performance of the cuBLAS batched \code{sgemm} in single precision on CUDA cores with the performance of a simple implementation of batched \code{GEMM} using CUDA~9 WMMA on Tensor Cores. Fig.~\ref{performance-batched} shows a box plot of the batched \code{GEMM} performance with and without Tensor Cores in white and grey boxes. The number of 16$\times$16 matrix multiplies, or \emph{batch size}, is represented on the $x$ axis of the plot, while the performance in Tflops/s is shown on the $y$ axis. The measured peak performance is 4~Tflops/s for 262,144 matrix multiplications with half precision input on Tensor Cores.
Increasing the number of 16$\times$16 matrix multiplies increases the performance of the \code{GEMM} with and without Tensor Cores. When using cuBLAS batched \code{sgemm} for $batch size > 131,072$, the system runs out of memory. For this reason, results for cuBLAS batched \code{sgemm} for $batch size > 131,072$ are not reported in the plot. The performance of our naive implementation of batched \code{GEMM} with half precision inputs outperforms the cuBLAS batched \code{sgemm} in full single precision. The performance of batched \code{GEMM} varies between 2.5$\times$ and 12$\times$ the performance of cuBLAS batched \code{sgemm} varying the \emph{batch size}.  
%We note the abrupt performance improvement of cuBLAS batched \code{sgemm} for \emph{batch size \textgreater 32,768}.
%On Tensor Cores, the naive implementation of batched sgemm always outperforms the sgemm by a factor of 2.5.

\subsection{Precision and refinement}
\label{performance-precision-refinement}
\begin{figure}[t!]
\begin{center}
\includegraphics[width=0.6\columnwidth]{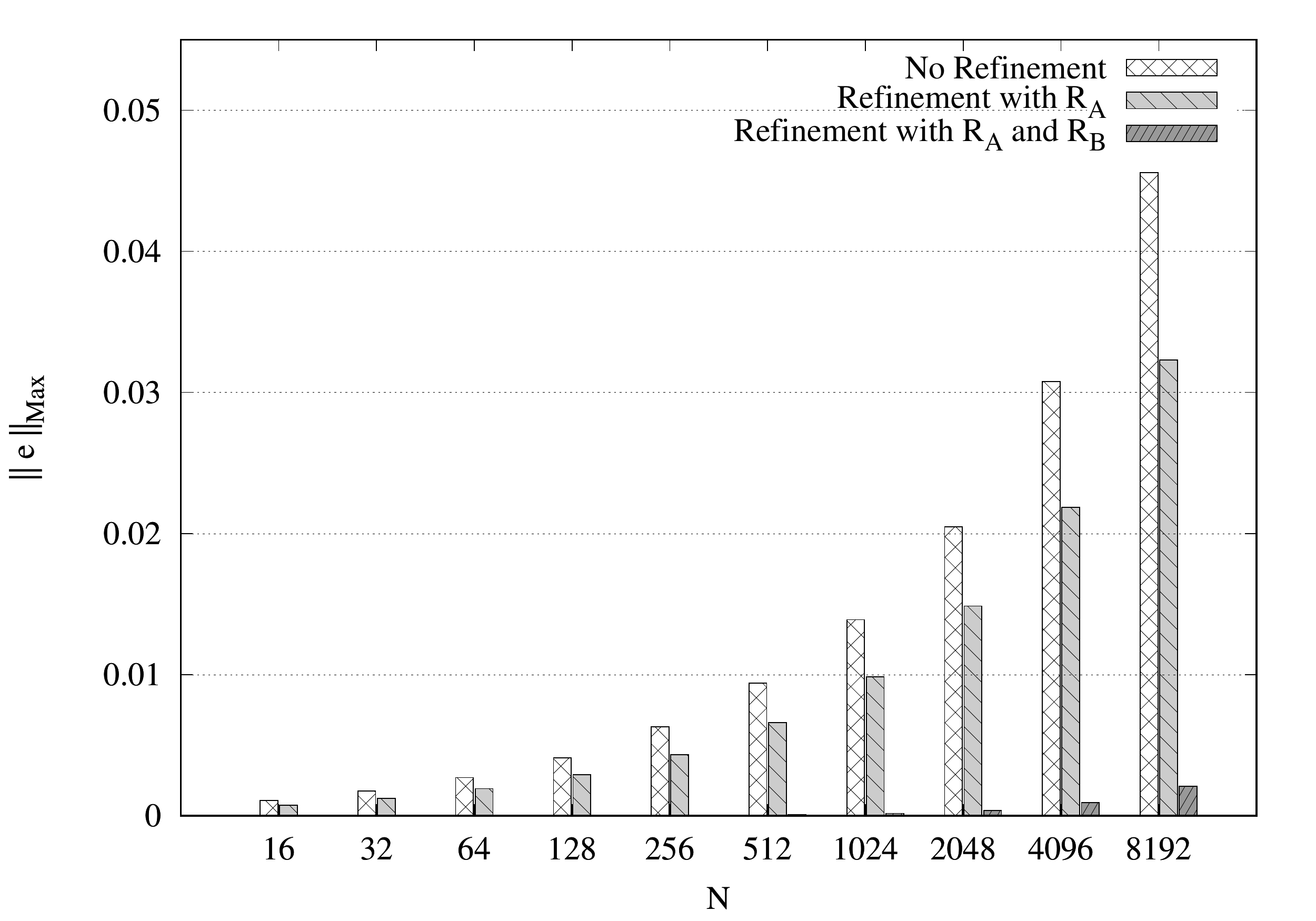}
\caption{Error in half precision (no refinement, white bars), using precision refinement with $R_A$, and precision refinement with both $R_A$ and $R_B$ varying the matrix size $N$.}%
\label{precision-loss}%
\end{center}
\end{figure}

We first measure the precision loss by half precision input on Tensor Cores. We then use Eqs. \ref{eq1} and \ref{eq2} to quantify the decrease of precision loss. Fig.~\ref{precision-loss} shows the error $\Vert e \Vert_{Max}$ for multiplications on Tensor Cores (white bars) varying with matrix size. Using the techniques in Eq. \ref{eq1} (light gray bars) and Eq. \ref{eq2} (dark grey bars), the precision loss can be decreased.
It is clear from Fig.~\ref{precision-loss} that by increasing the matrix size $N$, rounding error increases. This is due to the fact that the number of multiplications and summations for calculating one matrix element scales as $N^2$.  So, the error scales quadratically with $N$.

From Fig.~\ref{precision-loss} we can see that the use of Eq.~\ref{eq1} is only partially beneficial: we observe a 30\% decrease of the error for $N=8,192$. This small error decrease is due the fact that the norm of the two matrices is approximately the same. The use of Eq.~\ref{eq2} is more effective in decreasing the precision: the error is decreased by a factor of ten for $N=8,192$. We note that the precision loss strongly depends on matrix input values. For instance, if the $A$ and $B$ values are chosen randomly between $\pm$16 and $N=4,096$, we measure $\Vert e \Vert_{Max} = 8.32$ for $AB$ with no refinement, and $\Vert e \Vert_{Max} = 0.24$ for $AB$ with $A$ and $B$ refinement (Eq.~\ref{eq2}). In this case, the use of the refinement leads to a 35$\times$ decrease of error.

\begin{figure}[t!]
\begin{center}
\includegraphics[width=0.6\columnwidth]{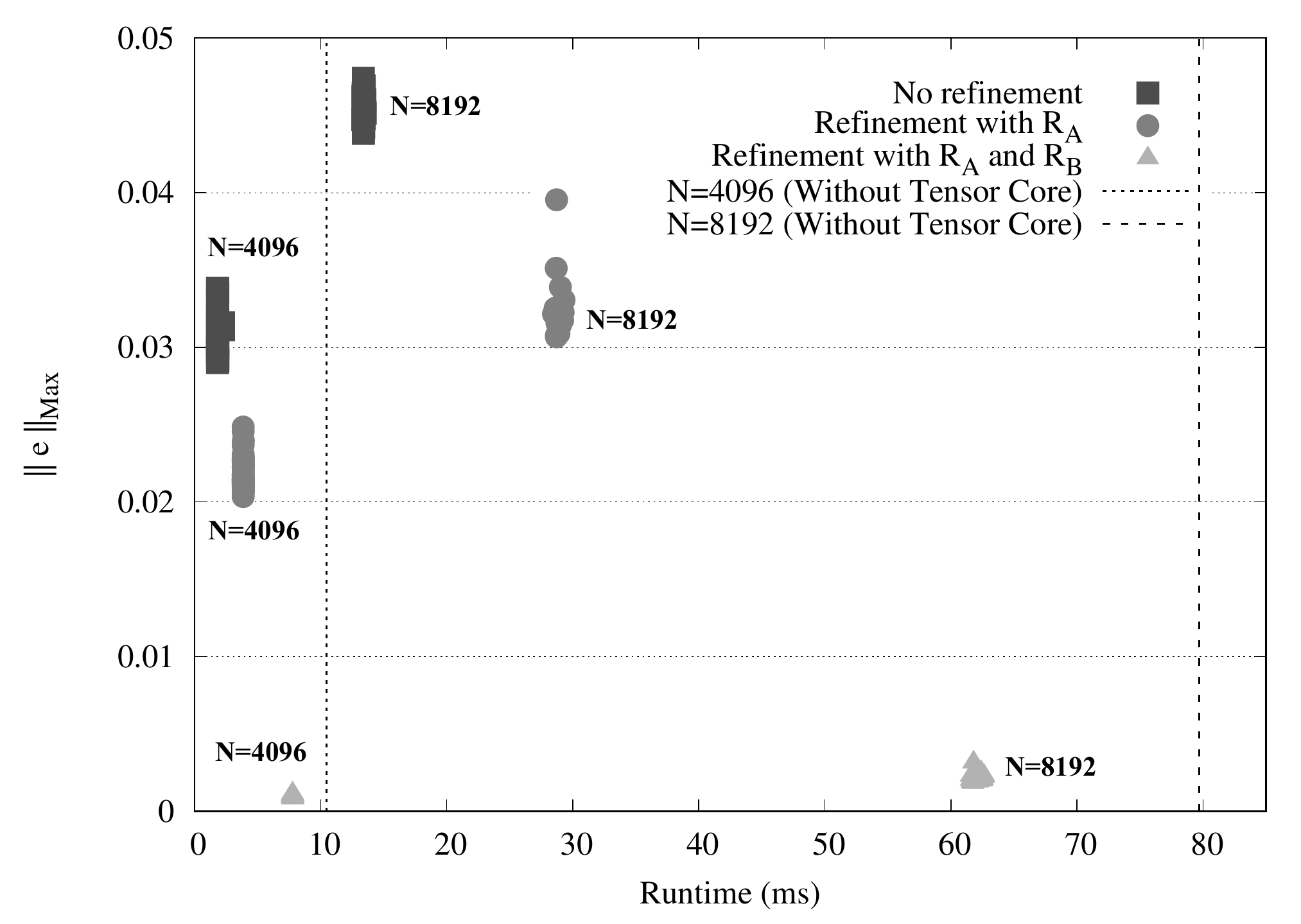}
\caption{Scatter plot with $\Vert e \Vert_{Max}$ on $x$ axis and runtime on the $y$ axis for \code{GEMM} with no refinement (squares), refinement with only $R_{A}$ (circles), and with both $R_{A}$ and $R_{B}$ (triangles) for $N=8,192$ and $N=4,096$. The two dashed lines represent the execution time for \code{sgemm} without Tensor Cores.}%
\label{runtime-refinement-offset}%
\end{center}
\end{figure}

Finally, we quantify the computational cost of applying the refinement technique discussed in Section \ref{precision} to decrease the precision loss when using NVIDIA Tensor Cores. Fig.~\ref{runtime-refinement-offset} presents a scatter plot in the execution time vs error plane for 8,192$\times$8,192 and 4,096$\times$4,096 matrix multiplication on Tensor Cores (square symbols), using precision refinement with $R_A$ (circle symbols) and with both $R_A$ and $R_B$ (triangle symbols). The scatter plot points are spread in the \emph{error} direction because we use random input values uniformly distributed between minus one and one as matrix entries. On the other hand, execution time measurements show little variation. In addition to scatter plot points, we add two lines at 10 and 80 ms to represent the average execution time recorded to perform a matrix multiplication in full single precision for $N=4,096$ and $N=8,192$, for which the error $\Vert e \Vert_{Max} $ is zero.

It is clear from Fig.~\ref{runtime-refinement-offset} that by increasing the computational cost (execution time) the error decreases. For $N=8,192$, if we increase the computational cost of a factor of 2.25, we can obtain a reduction of precision loss of approximately 30\% using precision refinement only with $R_A$. The precision refinement with both $R_A$ and $R_B$ leads to an error that is approximately 10$\times$ smaller than initial error with a 5$\times$ computational cost. The computational cost of precision refinement with $R_A$ and $R_B$ is still approximately 25\% lower than the cost of completing a \code{GEMM} without Tensor Cores. We note the implementation of precision refinement using four pipelined \code{GEMM} on Tensor Cores shown in \ref{pipeline} is not optimized as the precision refinement takes more than four times the time of completing one \code{GEMM}. For this reason, there is room for a large performance improvement.

%As seen in Figure \ref{precision-loss}, we can decrease the error by a factor of ten; however, this requires a runtime that is X times than calculating cuBLAS sgemm. We present the error and respective runtime of all tests being run for every $N$ in Figure \ref{runtime-refinement-offset}. We can clearly see the trade off between performance and precision with and without our refinement schemes.
%
%

\section{Discussion and Conclusions}
\label{discussion}
The NVIDIA V100 GPUs will be an important asset for the upcoming supercomputers, providing a large fraction of their overall computing power. The Volta microarchitecture features for the first time Tensor Cores, specially designed to perform tensor operations. We showed that the use of NVIDIA Tensor cores can boost the \code{GEMM} performance by 6$\times$ when multiplying large matrices and 2.5$\times$-12$\times$ when multiplying many small-size matrices in parallel. 
% The performance boost comes also at the cost of precision loss.

Many HPC applications are based on the multiplication of large matrices or several small-size matrix in parallel. For this reason, such applications can take direct advantage of NVIDIA Tensor Cores. On the other hand, some HPC applications, such particle-based codes, might require a reformulation of the algorithms in tensorial form to use the Tensor Cores effectively. In addition, a more in-depth study of the impact of mixed precision calculation on the overall simulation accuracy in large HPC applications is required to promote the uptake of NVIDIA Tensor cores by HPC applications.

In this paper, we focused on three main aspects when using NVIDIA Tensor Cores in HPC applications: programmability, performance and precision loss. We summarize our findings for each aspect as follows.
\begin{itemize}
\item {\bf Programmability.} Currently, there are three programming interfaces for developing applications using matrix-multiply-and-accumulate on NVIDIA Tensor Cores. CUDA~9 WMMA API provides direct access to CUDA Tensor Cores and can be used in combination. However, WMMA is a preview feature and will likely be modified in future releases. The other two ways of programming NVIDIA Tensor Cores are via CUTLASS and cuBLAS libraries. The CUTLASS implementation is based on WMMA and provides different tiling sizes that can be used for performance tuning. The NVIDIA cuBLAS library allows the use of Tensor Cores by setting cuBLAS math mode to \code{CUBLAS\_tensorOp\_MATH}. In this work, we have not covered support for convolution operations on Tensor Cores by the NVIDIA cuDNN~\cite{chetlur2014cudnn}, a library of primitives for deep neural networks because we focus on HPC usage of Tensor Cores. However, many of the concepts we covered in this paper can also be applied to cuDNN.

Finally, we note that from a high-level point of view, NVIDIA Tensor Cores can be seen as accelerators within an accelerator. Tensor Cores not only add considerable computing performance boost (a factor seven for \code{GEMM}), they also work on low precision and have their own local memory consisting of \emph{fragments}. When designing future interfaces for NVIDIA Tensor Cores, one possibility would be to treat Tensor Cores as accelerator-in-accelerator and investigate the use of direct kernel launching on NVIDIA tensor cores from the host. 

\item {\bf Performance.} We achieved the maximum performance in our test environment at 83~Tflops/s with cuBLAS \code{GEMM}. The naive \code{GEMM} implementation with CUDA WMMA did not lead to any performance improvement; however when using the implementation with CUDA \code{shared} memory, we measured a 5$\times$ performance improvement with respect to the \code{sgemm} performance on CUDA cores (not shown here). The problem size for maximum performance with \code{GEMM} is $N=8,192$. We also implemented a batched \code{GEMM} for Tensor Cores using CUDA~9 WMMA to evaluate the potential benefit of solving several small matrix multiplications in parallel. Although the implementation is not optimized, NVIDIA Tensor Cores still provided a performance increase of 2.5$\times$ - 12$\times$ with respect to performance of the cuBLAS batched \code{sgemm}. This shows that NVIDIA Tensor Cores could also be used to perform small-size matrix multiplications efficiently.

When investigating possible performance optimization, we noted that memory traffic still has high impact on the overall performance of the matrix multiplications, despite Volta's integration of L1 data cache and shared memory subsystem. For this reason, application optimization for NVIDIA tensor cores is likely to include strategies for data placement on the GPU memory subsystem.
%In addition, we found that the rounding operation to prepare the input in half precision takes a non-negligible computational cost, approximately 10\% of a \code{GEMM} operation on Tensor Cor.

An additional optimization is to use CUDA cores and Tensor Cores concurrently. This can be achieved by using CUDA streams in combination with CUDA~9 WMMA. This will also allow for more advanced and optimized pipelined mixed precision refinement methods and implementations.

\item {\bf Precision.} As matrix multiplication inputs are in half precision, precision loss occurs when they are rounded from single to half precision. In particular, precision loss is considerable when using large input values. When input matrix size increases, error increases because $\mathcal{O}(N^2)$ operations are required to calculate a matrix entry in the matrix multiplication. We showed a simple method to decrease precision loss by taking into account the rounding error when converting values from single to half precision. This method reduces the precision loss at the cost of increased computation. Further methods for increasing the precision can be developed, possibly taking advantage of single precision computation of unused CUDA cores while performing tensor operations.
\end{itemize}

In conclusion, despite the Volta microarchitecture with Tensor Cores has only been recently released, it is possible to program Tensor Cores for HPC applications using three approaches and achieve considerable performance boost at the cost of decreased precision. We expect the programming interfaces for NVIDIA Tensor Cores to evolve and allow increased expressiveness and performance. In particular, we noted that the measured Tensor Core maximum performance is still 74\% the theoretical peak, leaving room for improvement. While the precision loss due to mixed precision might be an obstacle for the uptake of NVIDIA Tensor Cores in HPC, we showed that it is possible to decrease it at the cost of increased computations. For all these reasons, it is very likely that HPC applications will strongly benefit from using of NVIDIA Tensor Cores. In the future we will focus on testing Tensor Cores in real-world HPC applications, such as Nek5000~\cite{offermans2016strong,markidis2015openacc} or Fast Multipole Method-accelerated FFT~\cite{cecka2017low}.

\section*{Acknowledgement}
This work used computing resources from KTH PDC Center for High Performance Computing. The Authors would like to thank Daniel Ahlin and Gilbert Netzer for their assistance when using the Tesla V100 at PDC. Funding for the work is received  from the European Commission H2020 program, Grant Agreement No. 671500 (SAGE). 
% Acknowledgement to sage %


\end{document}